# Magnetic structures on locally inverted interlayer coupling region of bilayer magnetic system

Chanki Lee,[1,2] Hee Young Kwon,[1] Nam Jun Kim,[1] Han Gyu Yoon,[1] Chiho Song,[1] Doo Bong Lee,[1] Jun Woo Choi,[2] Young-Woo Son,[3] and Changyeon Won[1, *]

[1]*Department of Physics, Kyung Hee University, Seoul 02447, South Korea*
[2]*Center for Spintronics, Korea Institute of Science and Technology, Seoul 02792, South Korea*
[3]*Center for Advanced Computation, Korea Institute for Advanced Study, Seoul 02455, South Korea*

We investigate the magnetic structures in a bilayer magnetic system with the locally inverted interlayer coupling region using Monte Carlo simulation. Stabilization of multiple magnetic structures including the magnetic skyrmion is possible in the locally inverted interlayer coupling region. Various factors such as the region area, anisotropy, interlayer coupling strength, and exchange coupling strength affects the properties of the structures including its size and chirality. We obtain conditions for their stabilization and for the magnetic structural transitions. Dzyaloshinskii-Moriya interaction (DMI) and the dipolar interaction play a prominent role as they enhance the formation and the stability of structures significantly. An asymmetric feature can arise from the broken inversion symmetry in the structure formation, and it gives an interfacial DMI, which stabilizes the skyrmion. It is realized that the dipole interaction also acts as an effective interfacial DMI in the system.

## I. INTRODUCTION

2D magnetic van der Waals (vdW) materials are in the spotlight as a further study of non-magnetic 2D vdW materials like few-layer graphene and h-BN [1-5, 9-11]. Moiré electronics [6,7] and moiré optics [8] emerged by twisting two of layers consisted with 2D vdW materials are also getting attention. The physical properties of twisted bilayer systems have been found to be deeply affected by rotational stacking fault between two layers. The combination of these two topics can lead to new topic named moiré magnetism, and studies for this topic are ongoing [9-11]. Accordingly, in twisted 2D magnetic vdW materials, the moiré pattern can expect to provide spatially varying interlayer coupling. It is also revealed that each regions of twisted bilayer $CrI_3$, $CrBr_3$ has ferromagnetic coupling and antiferromagnetic coupling by using first-principle calculation [9] and experiments [10]. The researches motivated us to introduce locally inverted interlayer coupling. On the other hand, the study on magnetic domains for the system is rarely conducted [11], thus more researches are necessary.

Two-dimensional (2D) magnetic structures have been actively studied recently due to their interesting physical properties [12-14] and possibilities in future applications [15,16]. Especially, the magnetic structures possessing a definite skyrmion number, such as magnetic skyrmions and magnetic vortices, have been extensively explored since they are known to be stable by topological protection [17], at the same being controllable with spin torques [18] or external fields [19]. The skyrmion number is a topological number defined by the number of spatial spin twist composing a magnetic structure. It is quantized; hence, not changed without topological phase transition that involves overcoming an energy barrier or symmetry breaking. A traditional method to generate magnetic chiral structures is to make structural edges by lithography so that the dipolar interaction around the edge builds rotational magnetic boundary [20]. This structure is commonly referred to as a vortex and its skyrmion number is either positive or negative half. Chiral magnetism also appears in magnetic systems with broken inversion symmetry in which the spin-spin interaction includes both the DMI and the exchange interaction [21]. Magnetic skyrmion with skyrmion number of either $+1$ or $-1$ can be stably generated in those systems.

In this paper, we propose a mechanism to generate magnetic structures of definite chiral numbers under locally varying interlayer coupling. The interlayer coupling between two magnetic layers has been used to control various magnetic properties [22-26]. When there is a local region of inverted interlayer coupling, multiple stable magnetic structures including magnetic skyrmion can be formed within the region. We investigate the formation condition of each magnetic structure by magnetic simulation. Not only do they have definite skyrmion numbers, but also a local structure resides in one of two coupled layers. The roles of the DMI and an additional effective DMI, which break the symmetries and the degeneracies of the systems as well as they contributing to the stabilization of the chiral structure, are investigated.

## II. MODEL AND COMPUTATIONAL METHODS

Figure 1(a) shows the model that we studied. Two ferromagnetic layers (layer 1 and layer 2) with perpendicular magnetic anisotropy are under interlayer exchange coupling that is locally inverted in a region of radius $R_0$. The magnetic energy is written as follows:

$$U_{tot} = -J\sum_{<i,j>}\vec{S}_i \cdot \vec{S}_j - K_z\sum_i|\vec{S}_{i,z}|^2 - J_{int}(r)\sum_i\vec{S}_{1,i}\cdot\vec{S}_{2,i} - \sum_{<i,j>}\vec{D}_{ij}\cdot(\vec{S}_i\times\vec{S}_j) - M^2\sum_{(i,j)}\frac{3(\vec{S}_i\cdot\vec{r}_{ij})(\vec{S}_j\cdot\vec{r}_{ij}) - r_{ij}^2(\vec{S}_i\cdot\vec{S}_j)}{r_{ij}^5} \quad (1)$$

where $\vec{S}$ is the normalized Heisenberg spin without dimension, $J$ is the exchange coupling strength in each layer, $K_z$ is the perpendicular magnetic anisotropy ($K_z > 0$), $J_{int}(r)$ is the interlayer coupling strength between the spins in the two layers ($\vec{S}_{1,i}$ in layer 1 and $\vec{S}_{2,i}$ in layer 2), $\vec{D}_{ij}$ is the DMI vector with dimension of energy per site, $M^2$ is the dipolar interaction strength, and $\vec{r}_{ij}$ is the displacement vector between $\vec{S}_i$ and $\vec{S}_j$. The summation for the exchange energy and DMI energy only includes interactions between the nearest neighborhoods in the same layer. We used a square grid model of a grid size $a$. The interlayer coupling strength, $J_{int}(r)$, is given by $-J_0\left[1-\left(\frac{r}{R_0}\right)^2\right]$, where $r$ is the radial distance from the center of each layer. The interlayer coupling changes its sign at $r = R_0$, and it favors anti-parallel alignment of spins inside the region and parallel alignment outside the region, when $J_0 > 0$. The coupling is inverted by changing the sign of $J_0$.

The magnetic structures stabilized in the system were estimated by theoretical considerations and confirmed by magnetic simulation. Initial structures used in the simulation were relaxed to find the energy minimum states. The spin vectors were iteratively calculated to follow their local effective fields until they did not vary forming stable states. The effective field is determined by $-\frac{\partial U_{tot}}{\partial \vec{S}_i}$, which effectively includes all the effects from the exchange interaction, anisotropy, and other involving interactions. We used a Heisenberg model of 131,072 sites ($256a \times 256a \times 2a$) with a periodic boundary condition. The dipolar interaction was calculated within the simulation sites in case it was considered. In the simulation, we used a Monte Carlo method with heat bath algorithm [27]. $R_0$ is fixed at $64a$ in the simulation. We note that the results can be generalized when we normalized $K_z$, $J_{int}$, $D_{ij}$, and $M^2$ to $\frac{R_0^2}{Ja^2}K_z$, $\frac{R_0^2}{Ja^2}J_{int}$, $\frac{R_0}{Ja}D_{ij}$, and $\frac{R_0^2}{Ja^2}M^2$, respectively.

Figure 1(c-f) shows four states among various stabilized magnetic states built in the locally inverted interlayer coupling region. For convenience of discussion, each of them was named the N-state (Fig. 1(c)), I-state (Fig. 1(d)), 0-state (Fig. 1(e)), and S-state (Fig. 1(f)). The N-state denotes the case in which the magnetization of both layers are uniformly aligned along the z direction. The I-state denotes the case in which the two layers have opposite in-plane spin directions within the locally inverted interlayer coupling region. The 0-state denotes the magnetic structure that has an inverted magnetization core with uniformly aligned in-plane wall and has almost uniform out-of-plane magnetization in the other layer. The S-state denotes the magnetic structure that has an inverted magnetization core with chiral in-plane wall and has almost uniform out-of-plane magnetization in the other layer.

The 0-state does not have rotational structure; thus, its skyrmion number is zero. In the I-state and 0-state, the in-plane magnetization direction can be any direction in each layer with antiferromagnetic interlayer alignment. Similar to the formation of an inverted magnetization core of 0-state in the one of the layers, in the S-state, a skyrmion also forms in either one of the layers while maintaining an almost uniform out-of-plane magnetization in the other layer. The skyrmion denotes magnetic structure with skyrmion number of +1, and we use the skyrmion terminology for the case that either of two layers has the structure inside the locally inverted interlayer coupling region as expressed in the Fig. 1(f).

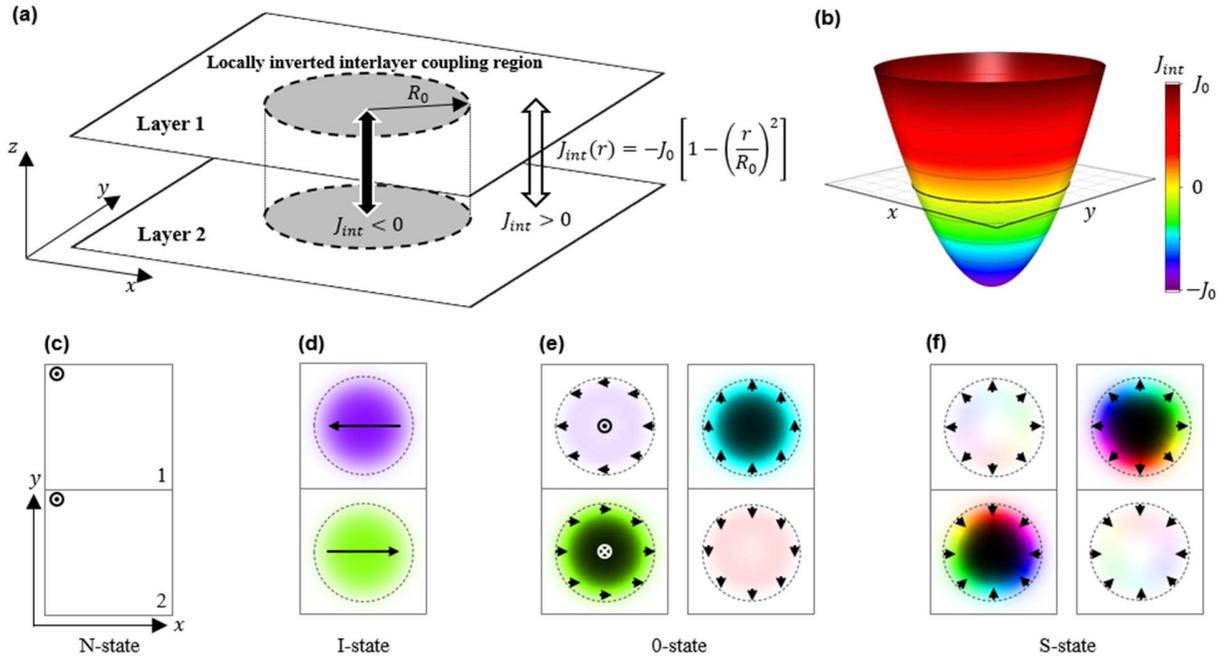

**FIG. 1. (a)** A schematic of bilayer magnetic system. The interlayer coupling inside the circular area of radius $R_0$ is assumed to be anti-parallel ($J_{int} < 0$) and the coupling outside is parallel ($J_{int} > 0$). It depends on $r$ in a quadratic way, the radial distance from a center of the coupling region as in **(b)**. **(c-f)** Stabilized magnetic structures possibly built in the region. The arrows and colors indicate the direction of the in-plane spin component, and brightness indicates the out-of-plane spin component. The left and right panels in (e) and (f) shows inverted magnetization core regions in layer 2 only and layer 1 only, respectively.

The competition among magnetic interactions determines which magnetic state is formed stably, and multiple states can exist simultaneously depending on the parameter conditions. For example, the N-state is built when the interlayer coupling is weak and the anisotropy is strong, and it can remain even if it is not the most energy-minimized state.

We explored the conditions for the local structures in Fig. 1(c-f) to be stabilized, using each initial structure of the simulation because the final magnetic state in the simulation would be affected by the initial spin configuration.

## III. RESULTS AND DISCUSSION

We first considered the cases where DMI and the dipolar interaction are negligible, i.e. only the first three terms in Eq. (1) are present in the simulation. The effect from DMI and the dipolar interaction will be discussed later. We let the given initial structures relax in the simulation and observed whether the structure was sustained or transformed to other structures for the given interaction parameters.

Figure 2(a) shows the final stable states in each perpendicular magnetic anisotropy and interlayer coupling strength when the initial structure is the 0-state. In the figure, the blue region indicates that the final structure remains as the 0-state, the white region indicates that the final structure is changed to the I-state, and the red region indicates that initial structure shrinks and annihilate to become the N-state. Therefore, the 0-state remained and was stabilized only in the blue region of Figure 2(a). If the anisotropy is weak or the size of the locally inverted interlayer coupling region is small ($K_z \frac{R_0^2}{a^2} \lesssim \frac{\pi^2}{4} J$), as shown with the yellow dashed line in Fig. 2(a) and (b), then the I-state (Fig. 1(d)) is favored. This can be explained in terms of the competition between the perpendicular magnetic anisotropy and interlayer coupling. The systems with stronger perpendicular magnetic anisotropy has narrower domain walls of a local structure with the wall width on the order of $\pi a \sqrt{\frac{J}{K_z}}$. The presence of interlayer coupling would result in a local structure with diameter $2R_0$, so that when the domain wall width of a local structure is larger than the diameter of the locally inverted interlayer coupling region ($\pi a \sqrt{\frac{J}{K_z}} \gtrsim 2R_0$), the I-state is favored.

The simulation results also show that the initial 0-state changes to N-state; thus, the local structure disappears in case the interlayer coupling strength is weak and anisotropy is large. The same result can be obtained when the size of the locally inverted interlayer coupling region is large.

Fig. 2(b) shows the final stable state when the initial structure in the simulation is the N-state. Compared to Fig. 2(a), the boundary between the blue region and the red region moved upward. Hence, in the region enclosed by the boundaries in Fig. 2(a) and (b), the 0-state and N-state are both stable states.

A phase transition between the different magnetic structures can be explained by the competition between the perpendicular magnetic anisotropy favoring the N-state and the interlayer coupling favoring to build a local structure. The energy cost to form a circular domain wall structure (domain wall of a local structure), such as those in the 0-state and S-state, and a local structure can be denoted approximately as

$$U_1 = 2\pi^2 aR\sqrt{JK_z} - 4\pi J_0 \left[\frac{R^2}{2} - \frac{R^4}{4R_0^2}\right]. \quad (2)$$

The first term $2\pi^2 aR\sqrt{JK_z}$ is the energy cost to make a domain wall of a local structure. The second term $-4\pi J_0 \left[\frac{R^2}{2} - \frac{R^4}{4R_0^2}\right]$ is the energy cost to make a local structure in the locally inverted interlayer coupling region. The competition between these two terms determines the formation of magnetic structures in Fig. 1(c-f). To minimize the energy cost, $\frac{\partial U_1}{\partial R}$ becomes zero and the condition gives $4\pi J_0 \left[R - \frac{R^3}{R_0^2}\right] = 2\pi^2 a\sqrt{JK_z}$ and $\frac{\partial^2 U_1}{\partial R^2}$ must be positive. From these conditions, the radius $R$ is determined and the value should be bigger than $\frac{R_0}{\sqrt{3}}$. If the radius of a local structure $R$ is smaller than $\frac{R_0}{\sqrt{3}}$, a local structure cannot be stabilized. Therefore, the critical radius for the structure is denoted by $R_c = \frac{R_0}{\sqrt{3}}$. By substituting $R$ by $R_c$ in the equation, $4\pi J_0 \left[R - \frac{R^3}{R_0^2}\right] = 2\pi^2 a\sqrt{JK_z}$, we obtain

$$\frac{J_0 R_0^2}{Ja^2} = \frac{3\sqrt{3}\pi}{4}\sqrt{\frac{K_z R_0^2}{Ja^2}}. \quad (3)$$

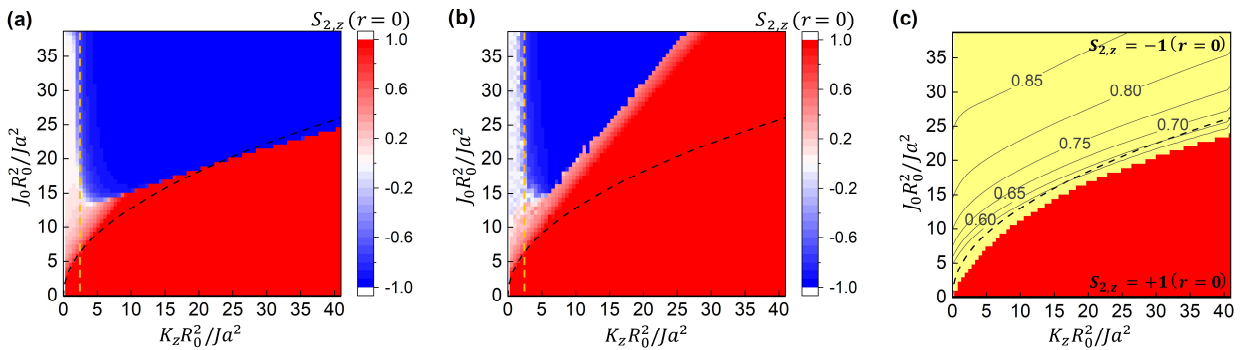

**FIG. 2. (a-c)** Phase diagrams obtained from the initial structure of the (a) 0-state, (b) N-state, and (c) S-state. The color represents the out-of-plane spin component at center of a local structure, $S_{2,z}$ in the center of the layer 2; hence, the blue, red, white and yellow colors indicate the final state is 0-state, N-state, I-state, and S-state, respectively. Each black contour curve has same $R/R_0$ in (c).

This equation gives the reason why a local structure is not obtained if $\frac{J_0 R_0^2}{Ja^2}$ is weaker than $\frac{3\sqrt{3}\pi}{4}\sqrt{\frac{K_z R_0^2}{Ja^2}}$, and the condition in Eq. (3) is drawn in the Fig. 2(a-c) with a black dashed line. Eq. (3) also provides the minimum size $R_0$ which is $\frac{R_0}{a} = \frac{3\sqrt{3}\pi}{4}\frac{\sqrt{K_z J}}{J_0}$. As $R_0$ increases, $J_0$ required to build the stable skyrmion structure decreases; thus, the structural phase transition can happen when $R_0$ is controllable as in the case of moiré superlattice of twisted 2D vdW materials. Such an inversely proportional $R_0 - J_0$ relation was also studied in ferromagnetic and antiferromagnetic coupling region in 2D magnetic vdW materials [11].

Fig. 2(c) shows the final stable state in each interlayer coupling strength and anisotropy when the initial structure in the simulation is the S-state. The S-state remains in the yellow region and disappears in the red region. Each black contour line drawn in the yellow region is for same radii expressed in $R/R_0$ of the skyrmion. The boundary between the yellow/red regions closely follows Eq. (3), similar to the blue/red boundary in Fig. 2(a); however, there is a discontinuous change of topological phase across the boundary in Fig 2(c), which is not the case for the boundaries in Fig. 2(a) and (b). The S-state has chiral structure that three stable magnetic structures of the 0-state, I-state, and N-state do not have. Therefore, the boundary is distinctive with a non-continuous change of $S_{2,z}(r = 0)$. The S-state does not appear in Fig. 2(a) and (b), since a topological phase transition from skyrmion number of 0 to that of +1 cannot occur with a non-chiral initial structure.

Fig. 3(a) shows $R/R_0$ of the local structure in the 0-state, and Fig. 3(b) shows $R/R_0$ of the skyrmion in the S-state as functions of interlayer coupling strength. Fig. 3(a) and 3(b) are for the cases in Fig. 2(a) and Fig. 2(c), respectively. The radius of the local structure changes drastically with certain interlayer coupling strength, which indicates the structural phase transition to the N-state. The condition for the phase transition varies with $K_z$; however, the radius at the drastic shrinking is mostly around $\frac{R_0}{\sqrt{3}}$ as expected. While the overall tendencies in Fig. 3(a) and (b) are similar, we find that the systems with smaller $K_z$ show larger difference in the radius (see black line for $\frac{K_z R_0^2}{Ja^2} = 4$ case in Fig. 3). For small $K_z$, if the initial structure is the 0-state, there is an additional phase, the I-state, between the initial 0-state and N-state, all these structures having zero skyrmion number; however, if the initial structure is the S-state, it undergoes a topological phase transition into the N-state directly. Therefore, the S-state is sustained longer than the 0-state.

We now consider the cases when DMI and the dipolar interaction are *not* negligible. As they play important roles in the formation of magnetic structure, it is necessary to investigate their roles in the evolution of the local structures that we have discussed.

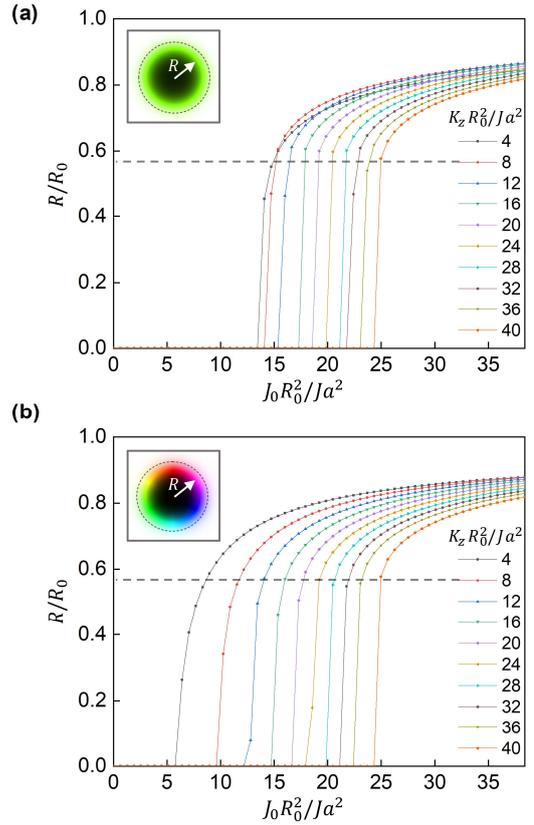

**FIG. 3.** (a), (b) $R/R_0$ as a function of interlayer coupling strength in each perpendicular magnetic anisotropy. The initial structure is the 0-state in (a) and the S-state in (b). The gray dashed line marks $R = \frac{R_0}{\sqrt{3}}$.

Fig. 4(a) shows the phase boundary of the S-state under the presence of DMI. We find that the phase boundary moves downward as the strength of DMI increases. In other words, the required interlayer coupling strength for a stable S-state decreases as the strength of DMI increases for the same anisotropy. Fig. 4(b) and (c) show $R/R_0$ of the structure as a function of interlayer coupling strength. Each curve in Fig. 4(b) is for different strength of DMI with fixed anisotropy, while those in Fig. 4(c) is for different strength of the anisotropy with fixed strength of DMI. Comparing Fig. 4(c) and Fig. 3(b), we find that the systems with low anisotropy ($\frac{K_z R_0^2}{Ja^2} \lesssim 20$) show drastic change with DMI: the S-state stabilized by DMI is sustained in the low coupling cases, whereas it would change into the N-state without DMI.

DMI gives an additional energy cost component, $-2\pi^2 aR|\vec{D}_{ij}|$, in Eq. (2) that stabilizes the S-state. The total energy with DMI is approximately,

$$U_2 = 2\pi^2 aR\sqrt{JK_z} - 2\pi^2 aR|\vec{D}_{ij}| - 4\pi J_0 \left[\frac{R^2}{2} - \frac{R^4}{4R_0^2}\right]. \quad (4)$$

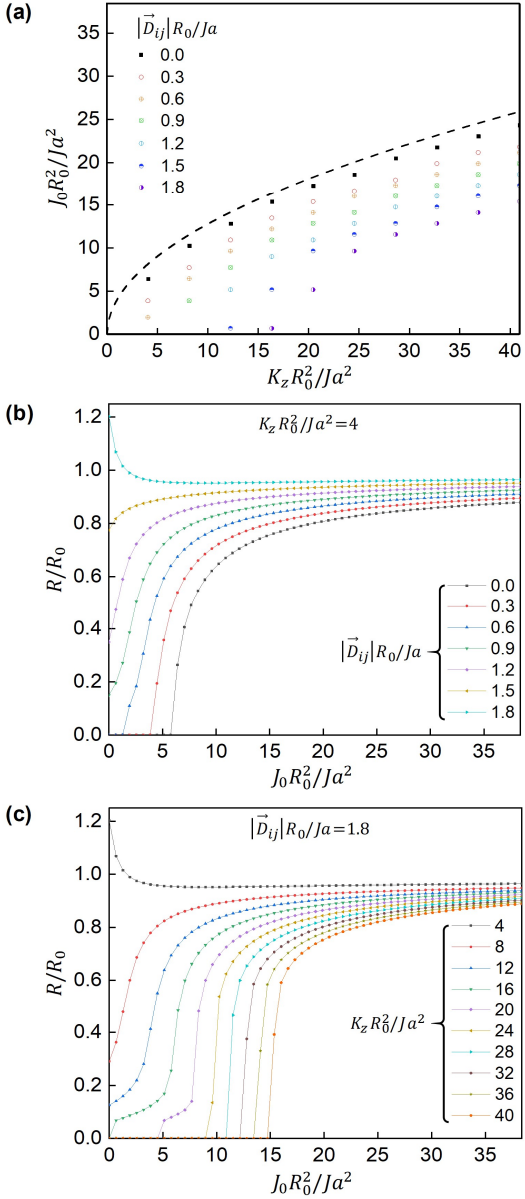

**FIG. 4. (a)** The phase boundary between the S-state and N-state for various DMI strengths. The top dashed line is from Eq. (3). The boundary gets lower in the diagram as DMI strength increases, which shows DMI stabilizes S-state. **(b-c)** $R/R_0$ as a function of interlayer coupling strength. The anisotropy is fixed in (b) and DMI strength is fixed in (c).

To minimize the energy cost, $\frac{\partial U_2}{\partial R}$ becomes zero and the condition gives $4\pi J_0 \left[R - \frac{R^3}{R_0^2}\right] = 2\pi^2 a\sqrt{JK_z} - 2\pi^2 a|\vec{D}_{ij}|$, with $\frac{\partial^2 U_2}{\partial R^2}$ being positive. Similar to the cases in Fig. 3(a) and (b), we introduce the concept of critical radius for the systems. By substituting $R$ by $R_c$ in the equation, we find

$$\frac{J_0 R_0^2}{Ja^2} = \frac{3\sqrt{3}\pi}{4}\left(\sqrt{\frac{K_z R_0^2}{Ja^2}} - \frac{|\vec{D}_{ij}|R_0}{Ja}\right). \quad (5)$$

This equation provides the critical conditions for the S-state to be sustained. If $\frac{J_0 R_0^2}{Ja^2}$ is lower than $\frac{3\sqrt{3}\pi}{4}\left(\sqrt{\frac{K_z R_0^2}{Ja^2}} - \frac{|\vec{D}_{ij}|R_0}{Ja}\right)$, the S-state becomes unstable and changes to the N-state. Compared to Eq. (3), the minimum requirement for interlayer coupling strength $J_0$ is reduced by the introduction of DMI. This explains why the S-state is more stable under DMI.

From the condition of $\frac{\partial U_2}{\partial R} = 0$, the size of local structures can be estimated, when $R$ is approximated to be close to $R_0$, we obtain

$$\frac{R}{a} = \frac{R_0}{a} - \frac{\pi}{4}\left(\frac{\sqrt{JK_z}}{J_0} - \frac{|\vec{D}_{ij}|}{J_0}\right). \quad (6)$$

Therefore, if DMI is strong ($|\vec{D}_{ij}| > \sqrt{JK_z}$), the radius of the skyrmion is larger than $R_0$, which is clearly seen in some results in Fig. 4(b) and (c).

We also studied the role of the dipolar interaction in the formation of the local structures. In contrast to the case of Fig. 4, in which we considered DMI with the negligible dipole interaction, we now consider the dipole interaction with negligible DMI. The dipolar interaction also stabilizes the I-state since it reduces the perpendicular magnetic anisotropy by providing shape anisotropy. The anisotropy $K_z$ in the Eq. (1-6) is replaced with the effective anisotropy $K_{z,\text{eff}} = K_z - 2\pi M^2$ and the range of the I-state in Fig. 2 is enlarged.

Fig. 5(a-d) shows how the dipolar interaction changes the structure of the S-state with Bloch-type skyrmion as the local structure. We find that the role of the dipolar interaction becomes complicated in the system, and it behaves as an effective interfacial DMI. As the dipolar interaction increases, the magnetization of the Bloch-type wall structure tilts more perpendicular to the wall to become a Néel-type wall structure. Fig. 5(e) is the case of the appearance of skyrmion in the layer 1 under the same condition of Fig. 5(d) (skyrmion in layer 2). Interestingly, the magnetization at the wall for two cases are opposite to each other, as it prefers a certain chirality depending on the layer. The magnetization in the layer 1 is influenced by the dipole field around structure wall in layer 2 (marked with red arrow in Fig. 5(h)) and it affects the magnetization on the wall through interlayer coupling. Thus, the structure in layer 2 has a preferred chirality. Consequently, the dipolar interaction functions as an asymmetric interaction contributing a chirality. It plays a role of an interfacial DMI.

Similar phenomenon in three-dimensional multilayer magnetic system was reported theoretically [28] and experimentally [29-31]. In the structure called three-dimensional skyrmion, Bloch-like skyrmions are in the middle of the film and Néel-like skyrmions are at the top and bottom surface of the film. This structure has been observed in Fe/Gd multilayers [29], FeGe nanodisk [30], and bulk Cu$_2$OSeO [31]. The composite only including top layer and bottom layer of the three-dimensional skyrmion is called Néel caps.

The mechanism of building the Néel-type structure here is similar to that of Néel caps. As the dipolar interaction increases, in-plane magnetization in two layers are coupled more strongly and it produces the effective interfacial DMI. We can identify that the size of in-plane components of the other layer increases as the dipolar interaction strength increases in Fig. 5(i).

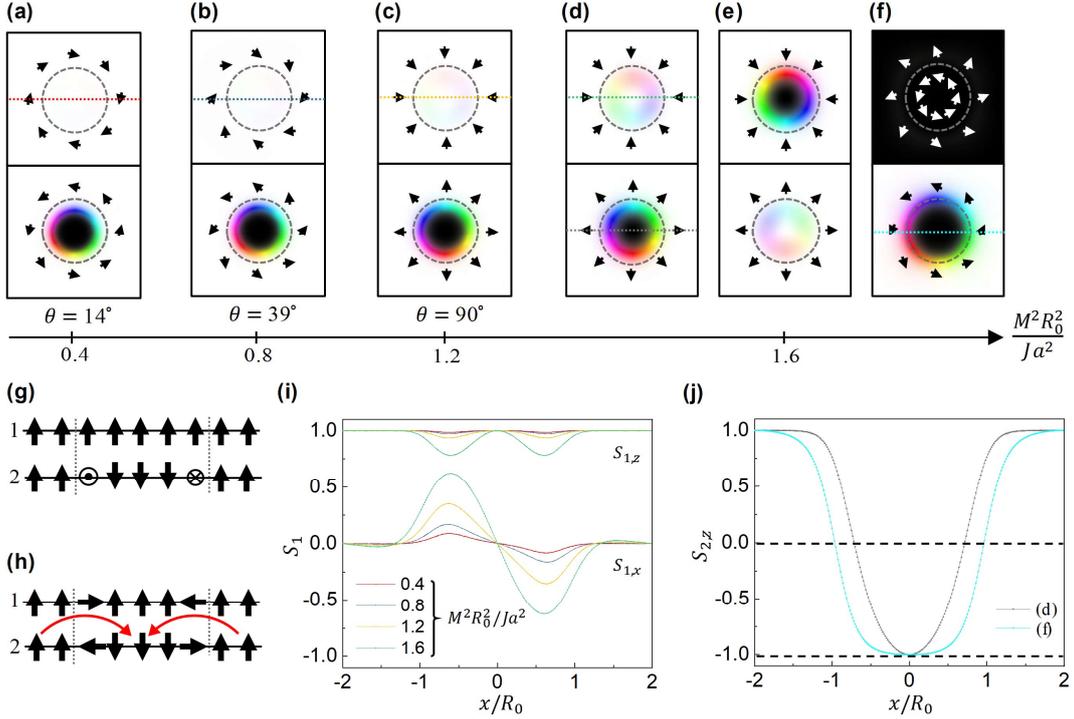

**FIG. 5.** **(a-e)** Spin configuration of top and bottom layer for several values of the dipolar interaction strengths. $\theta$ is in-plane spin components rotated. **(f)** Spin configuration of top and bottom layer in the case of reversed interlayer coupling ($J_0 < 0$). **(g)** A schematic spin configuration of initial state along x-direction in the middle of each layer **(h)** and that of state of (d). 1 and 2 mean layer 1 and layer 2. **(i)** The spin configuration of spin $x, z$ components for each $x$ position in the middle of top layer corresponding to the dotted lines in (a-d) **(j)** $S_z$ along the dotted lines in (d) and (f).

In Fig. 5(a-e), an inverted magnetization core of a skyrmion becomes smaller as the strength of the dipolar interaction increases; however, in case of $J_0 < 0$ (Fig. 5(f)), the size of an inverted magnetization core of a skyrmion is larger than in case of Fig. 5(d) and (e). Since the dipolar interaction favors parallel alignment of spins over the anti-parallel one, the outer region of parallel alignment increases with shrinking core region as shown in Fig. 5(d) and (e). If the interlayer coupling is reversed, the behavior is also reversed to increase the core region as shown in Fig. 5(f). In Fig. 5(j), the inverted magnetization core sizes of a skyrmion of Fig. 5(d) and (f) are compared.

To produce the local structures in experiments, we suggest $\frac{K_z R_0^2}{J a^2}$ of the experimental systems to be in the range from 0 to about 40. By adjusting the size of locally inverted interlayer coupling region $R_0$ and lattice constant of a crystal $a$, we can determine a ratio between $K_z$ and $J$. For example, the ratio is already studied in $CrCl_3$, $CrBr_3$, and $CrI_3$ [5,11]. Monolayer $CrBr_3$ has perpendicular magnetic anisotropy $K_z = 0.014J$. We can control the value of $R_0$ up to about 50~60 times of the lattice constant by changing the angle between two layers of $CrBr_3$, so that $\frac{K_z R_0^2}{J a^2}$ can be predicted to be in a range of 0~40. We also studied the magnetic states in the system with locally inverted interlayer coupling regions of a hexagonal array, which may be appeared in the twisted bilayer chromium tri-halides [32].

In addition, magnetic patterned array can be another candidate. In magnetic patterned array system, it is predicted theoretically that two-dimensional skyrmion lattice can appear [33], and it is already observed in practice [34]. If a magnetic patterned array is with locally varying interlayer coupling, the local structures expected in this study can be realized.

In summary, we have simulated for the bilayer magnetic system, which has locally inverted interlayer coupling region. We found that the various stable local structures can be generated and demonstrated how the competition among exchange interaction, perpendicular magnetic anisotropy, interlayer coupling affect the formation of stable magnetic states. We also identified the roles of DMI on the S-state and the dipolar interaction on the I-state and S-state.

## ACKNOWLEDGMENTS

This research was supported by Grants from the National Research Foundation (NRF) of Korea, funded by the Korean Government (NRF-2018R1D1A1B07047114). C. L. and J. W. C. acknowledges the KIST Institutional Program (2E29410), and the National Research Council of Science and Technology (NST) grant by MSIP (Grant No. CAP-16-01-KIST). Y.-W. S. was supported in part by NRF of Korea (Grant 2017R1A5A1014862, SRC program: vdWMRC Center).

# Supplemental Material for
# "Magnetic structures on locally inverted interlayer coupling region of bilayer magnetic system"

Chanki Lee,[1,2] Hee Young Kwon,[1] Nam Jun Kim,[1] Han Gyu Yoon,[1] Chiho Song,[1] Doo Bong Lee,[1] Jun Woo Choi,[2] Young-Woo Son,[3] and Changyeon Won[1, *]

[1]*Department of Physics, Kyung Hee University, Seoul 02447, South Korea*
[2]*Center for Spintronics, Korea Institute of Science and Technology, Seoul 02792, South Korea*
[3]*Center for Advanced Computation, Korea Institute for Advanced Study, Seoul 02455, South Korea*

**Simulation for the system with locally inverted interlayer coupling regions of a hexagonal array**

The moiré pattern made by two of hexagonal lattice has got great attention from researchers, especially for twisted bilayer graphene. Recently, a theoretical research performed using first principle calculation reveals that twisted bilayer graphene can be a ferromagnet [S1]. Many studies for the magnetism of 2D magnetic vdW materials of chromium tri-halides including $CrCl_3$, $CrBr_3$, and $CrI_3$ are also on the way because of their exotic magnetic properties [S2]. The chromium tri-halides have crystal structure of edge-sharing octahedral, which also can be arranged in a hexagonal lattice. Therefore, the research for the local structures of a hexagonal array within a bilayer magnetic system is also necessary. In an experimental way, if we made a bilayer system using a magnetic material, which has the hexagonal lattice when we observe the layers at the top, it would show the moiré pattern as changing the angle between two layers. Moreover, it is already revealed that bilayer chromium tri-halides can make locally inverted interlayer coupling regions of a hexagonal array from the first-principle calculation study [S3]. This can support our proposed magnetic system.

Here, to make locally inverted interlayer coupling regions of a hexagonal array, we gave different equation only for interlayer coupling in Hamiltonian as follows:

$$J_{int}(r) = J_1 \left[ \sin^2\left(\pi \frac{x}{64}\right) + \sin^2(-\pi \frac{x}{128} + \pi \frac{y}{128}\sqrt{3}) + \sin^2(-\pi \frac{x}{128} - \pi \frac{y}{128}\sqrt{3}) \right] - J_2 \quad \text{(S1)}$$

where the range of $x$ is from 1 to 256 and of $y$ is from 1 to 148. We fixed the value of $J_1$ at 0.2. Dipolar interaction is not considered.

Fig. S1. shows stabilized magnetic states in the system with locally inverted interlayer coupling region of hexagonal array. We notice that the magnetic state in Fig. S1(a), (b), and (m) corresponds to the I-state (Fig. 1(d)). In Fig. S1(n), total number of circles at the bottom layer is eight. There are seven 0-states (Fig. 1(e)) and one S-state (Fig. 1(f)) in the aspect of the local structure. Moreover, we obtained the triangular shaped local structures. The magnetic state in Fig. S1(h-l) and (v-x) is similar to the I-state, and in Fig. S1(u) is similar to the 0-state.

Unlike the magnetic system in Fig. S1, we consider DMI additionally for the system in Fig. S2. As a result, the hexagonal lattice consisted with skyrmions is obtained, which represented in Fig. S2(c) and (d). The system also gives skyrmion-like triangular structure (Fig. S2(e)). This structure is already reported in one of recent papers [S4]. Consideration for DMI makes the system more complex; thus, the spin configurations of Fig. S2(a), (b), and (f), which are not achieved in the system without DMI, are also gained.

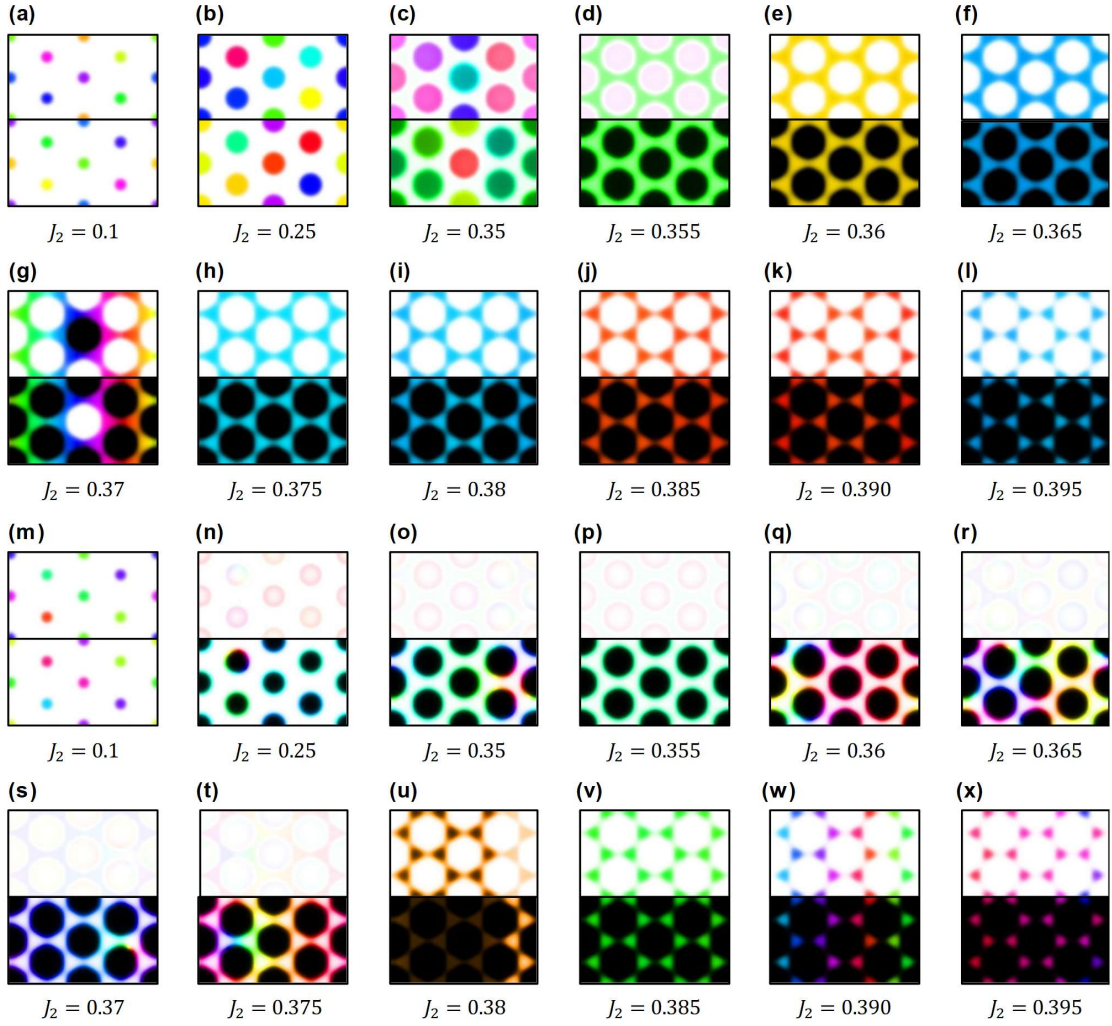

**FIG. S1. (a-x)** The local magnetic structures are in hexagonal array with (a-l) $K_z = 0.001$ or (m-x) $K_z = 0.010$. The parameter related to interlayer coupling, $J_2$, is written with its value below each spin configuration. DMI is not considered.

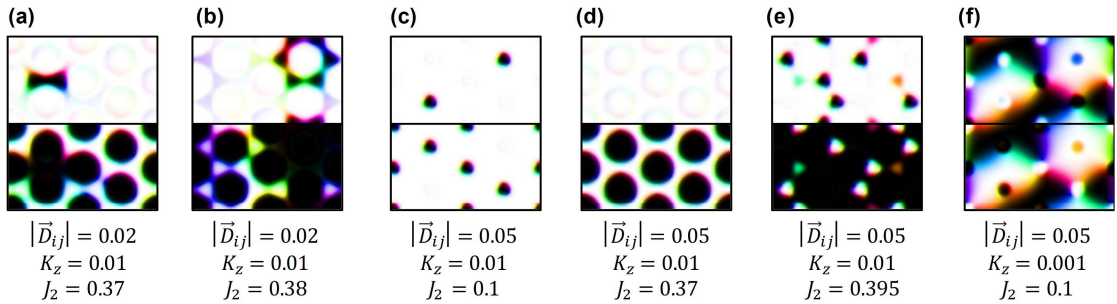

**FIG. S2. (a-f)** The local magnetic structures are in hexagonal array with each magnetic parameters. Parameters related to DMI, perpendicular magnetic anisotropy, and interlayer coupling are written with its value below each spin configuration.